\documentstyle[aps,amssymb,amsfonts,floats,epsfig]{revtex}

\newcommand{\bibepjb}{Eur. Phys. J. B }

\newcommand{\bmr}{\mbox{\boldmath $r$}}
\newcommand{\vp}{\mbox{\boldmath $p$}}
\newcommand{\vq}{\mbox{\boldmath $q$}}

\newcommand{\eqref}[1]{(\ref{#1})}

\newcommand{\sign}{{\rm sign}}

\newcommand{\lb}{\left(}
\newcommand{\rb}{\right)}

\begin{document}

\title{Effect of Angular momentum on equilibrium 
  properties of a self--gravitating system}

\author{Olivier Fliegans\footnote{Corresponding author, {\tt
      fliegans@hmi.de}}, D.H.E. Gross}

\address{Hahn--Meitner--Institut Berlin, Bereich Theoretische Physik
  (SF5), Glienicker Str. 100, D--14109 Berlin, Germany}

\date{\today}
\maketitle

\begin{abstract}
  The microcanonical properties of a two dimensional system of $N$
  classical particles interacting via a smoothed Newtonian potential
  as a function of the total energy $E$ and the total angular momentum
  $L$ are discussed. The two first moments of the distribution of the
  linear momentum of a given particle at a fixed position show that,
  (a) in average the system rotates like a solid body (b) the
  equipartition theorem has to be corrected by a term proportional to
  the square of the fluctuations of the inertial momentum of the
  system.  In order to estimate suitable observables a numerical
  method based on an importance sampling algorithm is presented. The
  entropy surface $S$ shows a negative specific heat region at fixed
  $L$ for all $L$. Observables probing the average mass distribution
  are used to understand the link between thermostatistical properties
  and the spatial distribution of particles.  In order to define a
  phase in non--extensive system we introduce a more general
  observable than the one proposed by Gross and Votyakov [\bibepjb
  {\bf 15}, 115 (2000)]. This observable is the sign of the largest
  eigenvalue of the entropy surface curvature. If it is negative then
  the system is in a pure phase; if it is positive then the system
  undergoes a first order phase transition. At large $E$ the
  gravitational system is in a homogeneous gas phase. At low $E$ there
  are several collapse phases; at $L=0$ there is a single cluster
  phase and for $L\neq0$ there are several phases with two clusters, the
  relative size of the clusters depends on $L$. All these pure phases
  are separated by first order phase transition regions. The signal of
  critical behaviour emerges at different points of the parameter
  space $\lb E,L \rb$. We also discuss the ensemble introduced in a
  recent pre--print by Klinko and Miller; this ensemble is the
  canonical analogue of the one at constant energy and constant
  angular momentum, i.e. it is defined at constant temperature and
  constant angular velocity. We show that a huge loss of informations
  appears if we treat the system as a function of intensive
  parameters: besides the known non--equivalence at first order phase
  transitions, there exit in the microcanonical ensemble some values
  of the temperature and the angular velocity for which the
  corresponding canonical ensemble {\em does not exist}, i.e. the
  partition sum diverges.
\end{abstract}


\section{Introduction}
\label{sec:introduction}
The thermostatistical properties of systems of $N$ classical particles
under a long--range attractive potential have been extensively studied since
the seminal work of
Antonov~\cite{antonov62},
\cite{padmanabhan90:_statis,1967MNRAS.136..101L,1968MNRAS.138..495L,hertel71:_solub_system_negat_specif_heat,thirring70:_system_negat_specif_heat,saslaw85:_gravit}.
One of their more specific and interesting properties is that they are
unstable for all $N$~\cite{padmanabhan90:_statis} and therefore not
thermodynamically extensive, i.e. they exhibit negative specific heat
regions even when the system is composed by a very large number of
particles.

It is quite natural to ask whether the total angular momentum $L$,
which is an integral of motion for systems of relevance in the
astrophysical context, plays a non--trivial role on the equilibrium
properties of these systems. Indeed $L$ is considered as an important
parameter in order to understand the physics of systems like
galaxies \cite{1999MNRAS.309..481L,lima00,binney87:_galac};
globular
clusters
\cite{lagoute96:_rotat_1,lagoute96:_rotat_2,horwitz77:_steep,klinko00:_mean_field_theor_spher_gravit_system,lynden-bell00:_rotat_statis};
molecular clouds in multifragmentation
regime \cite{combes98:_fract_struc_driven_self_gravit,vega:_fract_struc_scalin_laws_univer}
which might eventually lead to stellar
formation \cite{bate98:_collap_molec_cloud_core_stell_densit,klein98:_gravit_collap_fragm_molec_cloud,klessen97:_fragm_molec_cloud,1996MNRAS.280.1190B0,whitworth96:_star,white96:_violen_relax_hierar_clust,1990ApJ...363..197C}.

Previous works have already studied the effect of $L$ in the mean
field limit with a simplified potential and imposing a spherical
symmetry~\cite{klinko00:_mean_field_theor_spher_gravit_system,laliena99:_effec};
or at $L=0$~\cite{horwitz77:_steep}.

Our work, presented in this paper, is an attempt to overcome some of these
approximations.

Thermodynamical equilibrium does not exist for Newtonian
self--gravitating systems, due both to evaporation of stars (the
systems are not self--bounded) and short distance singularity in the
interaction potential. However there exists intermediate stages where
these two effects might be neglected and a quasi--equilibrium state
might be reached (dynamical issues like ergodicity, mixing or
``approach to
equilibrium''~\cite{saslaw85:_gravit,yawn97:_ergod,reidl93:_gravit}
are not considered in this paper).  In order to make the existence of
equilibrium configurations possible we have, first, to bound the
system in an artificial box and, second, to add a short distance
cutoff to the potential. The latter point can be seen as an attempt to
take into account the appearance of new physics at very short
distances (about the influence of this short cut
see~\cite{sommer-larsen97:_struc_isoth_self_gas_spher_soften_gravit,follana00:_therm_self_gravit_system_soften_poten}).
Another way to avoid the difficulties due to the short distance
singularity is to describe the function of distribution of the
``stars'' within a Fermi--Dirac statistic
\cite{1967MNRAS.136..101L,chavanis98:_system}.

The box breaks the
translational symmetry of the system, therefore the total linear
momentum $P$ and angular momentum $L$ are not conserved. Nevertheless
we assume that the equilibration time is smaller than the
characteristic time after which the box plays a significant
role~\cite{horwitz77:_steep,laliena99:_effec}. Therefore $P$ and $L$
are considered as (quasi--)conserved quantities. We put the center of
the box at the center of mass $R_{CM}$ which is also set to be the
center of the coordinates. Therefore $P=0$.

As already mentioned, self--gravitating systems are non--extensive and
a statistical description based on their intensive parameters should
be taken with caution since the different statistical ensembles are
only equivalent at the thermodynamical limit far from first order
transitions (see Sec.~\ref{sec:beta-beta-gamma}). Moreover, this limit
is required in order to define phases and phase transitions if one
fixes the intensive parameters~\cite{lee52:_statis_theor_II}.  In
contrast, the microcanonical ensemble (ME) does not require this
limit, it allows a classification of phase transitions for finite size
systems~\cite{gross00:_phase_trans_small,gross97:_microc}. Hence the
considered system is studied within the natural ME framework.

In order to perform the computation in a reasonable time we have to
consider a two dimensional system.

The paper is organized as follow: In section~\ref{sec:micr-prop} we
recall the analytical expressions for entropy and its derivatives
(Sec.~\ref{sec:micr-defin}), clarify the definition of phase
transitions for non--extensive systems proposed
in~\cite{gross00:_phase_trans_small}
(Sec.~\ref{sec:phase-phase-trans}), discuss the two first moments of
the distribution of the linear momentum of a given particle at a fixed
position (Sec.~\ref{sec:moment-aver-disp}), and present a numerical
method based on an importance sampling algorithm in order to estimate
suitable observables (Sec.~\ref{sec:numer-integr}). Numerical results
are presented in section~\ref{sec:results}; the link between the
average mass distribution and the thermostatistical properties is made
in Sec.~\ref{sec:mass-distribution}. In Sec.~\ref{sec:phase-diagram}
we use the definition of phase introduced in
Sec.~\ref{sec:phase-phase-trans} to draw the phase diagram of the
self--gravitating system as a function of its energy $E$ and angular
momentum $L$. We discuss the ensemble introduced in
\cite{klinko00:_mean_field_theor_spher_gravit_system}; therein this
ensemble is used to treat another model of rotating and
self--gravitating system. This ensemble is a function of the
(intensive) variables conjugate of $E$ and $L$. For our model we show
how the predictions using this ensemble are inaccurate and misleading
(Sec.~\ref{sec:beta-beta-gamma}). Results are summarized and discussed
in section~\ref{sec:comments-conclusions}.

\section{Microcanonical properties}
\label{sec:micr-prop}

\subsection{Microcanonical definitions}
\label{sec:micr-defin}

Consider a system of $N$ classical particles on a disk of radius $R$  
whose interaction is described by a Plummer softened
potential~\cite{plummer11,yepes97:_cosmol_numer}
\begin{equation}
  \label{eq:1}
  \varphi_{ij}=-\frac{Gm_im_j}{\sqrt{s^2+(\vq_i-\vq_j)^2}},
\end{equation}
where $m_i$ and $\vq_i=\{q_i^1,q_i^2\}$ are the mass and position of particle
$i$ respectively, $s$ is the softening length and $G$ is the gravitational
constant. The fixed total energy $E$ is described by the Hamiltonian
\begin{equation}
  \label{eq:2}
  {\cal H}=\sum_i\frac{\vp_i^2}{2m_i}+\varphi(\vq),
\end{equation}
where $\vp_i=\{p_i^1,p_i^2\}$ is the linear momentum of particle $i$,
$\varphi=\sum_{i<j}\varphi_{ij}$ and $\vq$ is a $2N$--dimensional vector whose
coordinates are $\{\vq_1,\ldots,\vq_N\}$, representing the spatial
configuration and is an element of the spatial configuration space
$V_c$, $\vq\in V_c\subset{\Bbb R}^{2N}$.

The entropy $S$ is given through the Boltzmann's principle (the Boltzmann
constant is set to 1)
\begin{equation}
  \label{eq:3}
   S(E,L,N)=\log\lb W(E,L,N) \rb,
\end{equation}
where $W(E,L,N)$ is the volume of the accessible phase--space at $E$,
$L$ and $N$ fixed (under the assumptions given in
\ref{sec:introduction})
\begin{eqnarray}
  \nonumber
  W\big(E,L,N\big)&=&\frac{1}{N!}\int\prod_{i=1}^N\left(
    \frac{d\vp_i\-d\vq_i}{(2\pi\hbar)^2} \right)
  \delta\big(E-{\cal H}\big)\delta^{(2)}\big(\sum_i\vp_i\big) \\
  \label{eq:4}
  & &   \times\delta\big(L-\sum_i\vq_i\times \vp_i\big) \delta^{(2)}\big(\sum_i\vq_i\big),
\end{eqnarray}
where $\vq_i\times \vp_i=q_i^1p_i^2-q_i^2p_i^1$.  After integration over
the momenta Eq.~\eqref{eq:4}
becomes~\cite{laliena99:_effec,calvo98:_monte_carlo}
\begin{equation}
  \label{eq:5}
  W\big(E,L,N\big)={\cal C}\int_{V_c}d\vq\:\frac{1}{\sqrt{I}}
  E_r^{N-5/2},
\end{equation}
where ${\cal C}=\frac{(2\pi)^{(N-3/2)}\prod_im_i}{(2\pi\hbar)^{2N}
  N!(\sum_im_i)\Gamma(N-3/2)}$ is a constant, $I=\sum_im_i\vq_i^2$ is the
inertial momentum and $E_r=E-\frac{L^2}{2I}-\varphi$ the remaining energy .
From the point of view of the remaining energy, if $L\neq0$ we can
already notice that the equilibrium properties are the results of a
competition between two terms; the rotational energy $\frac{L^2}{2I}$
and the potential energy $\varphi$. The former tries to drive the particles
away from the center of mass in order to increase $I$ whereas the
latter tries to group the particles together in order to decrease $\varphi$,
but since the center of mass is fixed this will lead to a
concentration of particles near the center and consequently will
decrease $I$.

The microcanonical temperature $T$ is defined by
\begin{equation}
  \label{eq:6}
    \frac{1}{T}=\beta\equiv\frac{\partial S}{\partial E}=\langle\frac{N-5/2}{E_r}\rangle,
\end{equation}
where $\langle\cdot\rangle$ is the microcanonical average
\begin{equation}
  \label{eq:7}
  \langle{\cal O}\rangle=\frac{{\cal C}}{W}\int_{Vc}d\bmr\frac{{\cal O}\left(\bmr\right)}{\sqrt{I}}
  E_r^{N-5/2}.
\end{equation}
The angular velocity $\omega$ is defined as minus the conjugate force
of $L$ times T~\cite{b.89:_physiq_statis}
\begin{equation}
  \label{eq:8}
  -\omega\equiv\frac{1}{\beta}\frac{\partial S}{\partial
    L}=-\frac{\langle\frac{L}{I}E_r^{-1}\rangle}{\langle E_r^{-1}\rangle}
\end{equation}

We also define $\gamma\beta$ has the conjugate of $L^2$:
\begin{eqnarray}
  \label{eq:9}
  \gamma\beta&\equiv&\frac{\partial S}{\partial L^2}=-\langle \frac{1}{2I}\frac{N-5/2}{E_r}\rangle , \\
  \label{eq:10}
  \omega&=&2L\gamma.
\end{eqnarray}

\subsection{Phase and phase transitions}
\label{sec:phase-phase-trans}

For small or self--gravitating systems a phase (or a phase transition)
can not be defined in the usual way using for example the Lee and Yang
singularities~\cite{lee52:_statis_theor_II} since these singularities
show up only at the thermodynamical limit. Invoking the
thermodynamical limit when studying small systems washses out all the
finite size effects that may lead to new phenomena, (e.g.
isomerisation of metallic clusters~\cite{kunz94:_multip},
multifragmentation of nuclei~\cite{gross95:_statit}) and for
self--gravitating systems the thermodynamical limit does not exist.
Hereafter, we will call ``Small systems'', those where the range of
the forces is of the order of the system size (e.g.  metallic
clusters, nuclei and self--gravitating systems) and also systems
without proper thermodynamical limit (e.g.  unstable
systems~\cite{posch90:_dynam}).

In a recent paper~\cite{gross00:_phase_trans_small} definitions for
pure phases and phase transitions (first and second kind) based
on the {\em local\/} topology of the microcanonical entropy surface
has been proposed. In the following we first fix some notations and
then recall the definitions.

Consider the microcanonical ensemble (ME) of an isolated physical
system. Its associated entropy $S(X)$ is a function of ${\cal N}$
``extensive'' dynamical conserved quantities $X=\{X^1,\ldots ,X^{\cal N}\}$.
Note that $X$ may not contain {\em all\/} the dynamical conserved
quantities and for simplicity all these parameters are considered as
being continuous. The Jacobian of $S(X=X_0)$ is noted by
$J_S(X_0)=\|\frac{\partial^2S}{\partial X^i\partial X^j}\|_{X_0}$, its eigenvalues are
$\{\lambda_1,\ldots,\lambda_{\cal N}\}$ where $\lambda_1\geq\lambda_2\geq\ldots\geq\lambda_{\cal N}$ and the determinant
of $J_S(X)$ is $D_S=\lambda1\cdots\lambda_{\cal N}$.

In~\cite{gross00:_phase_trans_small} phase transitions are defined
``{\em by the points and regions of non--negative curvature of the
  entropy surface [\ldots] as a function of the mechanical quantities\/}''.
Therein the sign of $D_S$ is put forward as a measure the concavity of
$S$ (its negative curvature) so that at first order phase transition
\begin{equation}
  \label{eq:23}
  \sign \lb D_S \rb =\sign \lb (-1)^{{\cal N}+1}\rb.  
\end{equation}
Though this condition is {\em necessary\/} it is not {\em
  sufficient\/} in the general case. In fact $S$ is a non--concave
function at $X_0$ if
\begin{equation}
  \label{eq:24}
  \lambda_1\geq0,  
\end{equation}
i.e. if at least one eigenvalue of $J_S$ is non--negative. Note that
in the two dimensional example model studied
in~\cite{gross00:_phase_trans_small} in order to illustrate the
definition (\ref{eq:23}), $\lambda_2$ is always negative therefore the sign
of $D_S$ is simply minus the one of $\lambda_1$, and the conditions
(\ref{eq:23}) and (\ref{eq:24}) are equivalent.

By using $\sign(\lambda_1)$, one can somewhat extend or clarify the
classification of phase transitions in non--extensive systems when the
entropy is a function of ${\cal N}\geq1$ variables for any set of the
parameters $X_0$
\begin{itemize}
\item A single pure phase if $\lambda_1(X_0)<0$.
\item A first order phase transition if $\lambda_1(X_0)>0$. As mentioned
  in~\cite{gross00:_phase_trans_small,gross97:_fragm_i} the depth of
  the associated entropy intruder is a measure of the intra--phase
  surface tension; how to define and to measure these depth when
  ${\cal N}>1$ will be discussed elsewhere. Note that some eigenvalues
  might still be negative although $\lambda_1(X_0)>0$ just like in the model
  presented in~\cite{gross00:_phase_trans_small}. In this case
  ``good'' order parameters are linear combinations of the
  eigenvectors whose eigenvalues are positive.
\item If $\lambda_1(X_0)=0$ and $\lambda_1$ is the only zero eigenvalue and
  $\nabla_{{\bf v}_1}\lambda_1=0$, where ${\bf v}_1$ is the eigenvector
  associated with $\lambda_1$ then there is a second order phase transition
  at $X_0$.
\item If several eigenvalues obey $\lambda_i=0$ and $\nabla_{{\bf v}_i}\lambda_i=0$ for
  $i=1,\ldots,n\leq{\cal N}$ then $X_0$ is a multicritical point.
\end{itemize}

\subsection{Momentum average and dispersion}
\label{sec:moment-aver-disp}

In this section we derive the average and the dispersion of the linear
momentum of a particle, we also compute its mean angular velocity
and relate it to the one of the system as defined in Eq.~(\ref{eq:8}).

The derivation of $\langle \vp_k\rangle_{\vq_k}$ the average momentum of particle
$k$ at fixed position $\vq_k$ (while the other
particles are free) is similar to that of $W$. Details of
the derivation can be found in Appendix~\ref{sec:appendix}, and the
result is 
\begin{equation}
  \label{eq:11}
  \langle \vp_k\rangle_{\vq_k}=L\langle I^{-1}\rangle_{\vq_k}m_k\sum_{\alpha,\delta=1}^2\epsilon_{\alpha\delta}q_k^\delta{\bf e}_\alpha,
\end{equation}
where ${\bf \epsilon}$ is the antisymmetric tensor of rank 2 and ${\bf e}_\alpha$ the
unit vector of coordinate $\alpha$. Equation~(\ref{eq:11}) shows that 
$\langle\vp_k\rangle_{\vq_k}$ is a vector perpendicular to $\vq_k$ whose module
is a function of $\|\vq_k\|$. In other words the orbit of a particle is
in the mean circular (this result is expected since the system is
rotationally symmetric). One can compute $\langle \omega_k\rangle_{\vq_k}$ the mean
angular velocity of $k$ at distance $\|\vq_k\|$ by first considering
$\langle L_k\rangle_{\vq_k}$ the mean angular momentum of $k$ at distance $\|\vq_k\|$
\begin{eqnarray}
  \nonumber
  \langle L_k\rangle_{\vq_k}&\equiv&\vq_k \times \langle\vp_k\rangle_{\vq_k} \\
  \label{eq:12}
  &=& L\langle I^{-1}\rangle_{\vq_k} I_k,
\end{eqnarray}
where $I_k=m_k\vq_k^2$.
The angular mean velocity of a particle on a circular orbit is classically
linked to $\langle L_k\rangle_{\vq_k}$ by
\begin{equation}
  \label{eq:13}
  \langle L_k\rangle_{\vq_k} = \langle\omega_k\rangle_{\vq_k} I_k.
\end{equation}
We can identify $\langle\omega_k\rangle_{\vq_k}$ in~(\ref{eq:12}) as
\begin{equation}
  \label{eq:14}
  \langle\omega_k\rangle_{\vq_k}=L\langle I^{-1}\rangle_{\vq_k}.
\end{equation}
The dependence of $\langle\omega\rangle_{\vq_k}$ on $||\vq_k||$ is of the order $1/N$
($\langle I^{-1}\rangle_{\vq_k}=\langle I^{-1}\rangle+{\cal O}(N^{-1})$),
therefore for large $N$ we can write (see Eq.~(\ref{eq:8}))
\begin{equation}
  \label{eq:15}
  \langle\omega_k\rangle_{\vq_k} \sim L\langle I^{-1}\rangle\approx\langle\omega\rangle.
\end{equation}
For large $N$ the mean angular velocity is the same for all the
particles at any distance from the center, in other words the system
in the mean rotates like a solid body. Moreover $\langle\omega_k\rangle_{\vq_k}$
corresponds to the thermostatistical angular velocity $\omega$ defined by
Eq.~(\ref{eq:8}). These are also classical results for extensive
systems at low $L$\cite{b.89:_physiq_statis,landau94:_physiq}. Note
also that these results do not depend on the interaction potential
$\varphi$.

The momentum dispersion $\sigma_{\vp_k}$ can also be derived. Using
Eq.~(\ref{eq:11}) and Eq.~(\ref{eq:36}), we get for large $N$
\begin{eqnarray}
  \nonumber
  \sigma_{\vp_k}^2 &\equiv & \langle\vp_k^2\rangle_{\vq_k}-\langle\vp_k\rangle^2_{\vq_k}\\
  \label{eq:16}
  &\sim &2\frac{m_k}{\beta}+I_kL^2m_k \lb \langle I^{-2}\rangle-\langle I^{-1}\rangle^2 \rb 
\end{eqnarray}
The second term of Eq.~(\ref{eq:16}) is proportional to the square of
the dispersion of $I^{-1}$ and to $\vq_k^2$ ($I_k=m_k\vq_k^2$). When
this term vanishes relatively to the first one, e.g. when the
fluctuations of $I^{-1}$ are small, or at high energy (low $\beta$) and
low $L$, we recover the usual dispersion of the Maxwell--Boltzmann
distribution. This term also gives a correction to the usual
equipartition theorem; for large $N$
\begin{eqnarray}
  \nonumber
  \langle E_k\rangle&\equiv&\frac{\sigma_{\vp_k}^2}{2m_k}\\
  \label{eq:17}
  &\sim&T+\frac{I_kL^2}{2} \lb \langle I^{-2}\rangle-\langle I^{-1}\rangle^2 \rb,
\end{eqnarray}
where $\langle E_k\rangle$ is the average internal kinetic energy (without the
contribution from the collective rotational movement) of particle $k$.
Again this correction is position--dependent via $I_k$.  In the
regimes where the fluctuations of the mass distribution can not be
neglected in Eqs.~(\ref{eq:16}) and (\ref{eq:17}) an estimation of the
temperature based on the velocity dispersion would show that the
temperature is smaller in the core than at the edge.
 
\subsection{Numerical method}
\label{sec:numer-integr}

From now on we set $m_i=m,\: \forall i=1,\cdots,N$ and use the following
dimensionless variables: (i) $E \to \epsilon=\frac{ER}{Gm^2N^2}$; (ii) $L \to
l^2=\Omega=\frac{L^2}{2Gm^3RN^2}$; (iii) $s \to \sigma=\frac{s}{R}$; (iv) $q \to
r=\frac{q}{R}$; (v) $V_c \to v_c$; (vi) $\varphi \to \phi = \frac{R}{Gm^2N^2}\varphi =
\frac{-1}{N^2}\sum_{i<j}\frac{1}{\sqrt{\sigma^2+(\bmr_i-\bmr_j)^2}}$; (vii) $I
\to I=\sum_ir_i^2$. The weight is now
  
\begin{equation}
  \label{eq:18}
  W\left(\epsilon,\Omega\right)={\cal C'}\int_{v_c}d\bmr\frac{1}{\sqrt{I}} \epsilon_r^{N-5/2},
\end{equation}
where $\epsilon_r=\epsilon-\frac{\Omega}{I}-\phi$ is the dimensionles remaining energy and
${\cal C'}$ a constant. Later on we no longer write this constant
since it plays no significant role (it only shifts the entropy by
$\log{{\cal C'}}$). The derivatives of entropy ($\beta$, $\omega$, \ldots ) are now
dimensionless quantities.

One usually estimates \eqref{eq:18} by means of some Monte Carlo
algorithm, updating the positions ${\bf q}$ by some small amount
$\delta{\bf q}$ in order to get a good pass acceptance and using the
configuration weight $W(\bmr)=\frac{1}{\sqrt{I}} \epsilon_r^{N-5/2}$ in the
Metropolis pass. Unfortunately this strategy does not really work
(within a reasonable \texttt{CPU}--time), because the $2N$--dim
configuration weight--landscape at fixed $\epsilon$ and $\Omega$ presents troughs
and high peaks~\cite{torcini99:_equil_n}, so exploring the total
configuration--space (or at least a subset containing the highest
peaks) would take a very long, in practice infinite, time. This
weight--landscape looks like the energy--landscape found in
spin--glass systems.

The strategy we have adopted is described in the following. First we
can rewrite \eqref{eq:18}
\begin{equation}
  \label{eq:19}
   W\left(\epsilon,\Omega\right)=\int dI\:d\phi\: Bg \lb I,\phi\rb \frac{1}{\sqrt{I}}
  \epsilon_r^{N-5/2},
\end{equation}
where $Bg \lb I,\phi\rb = \int_{v_c} d\bmr \delta\lb I'(\bmr)-I\rb \delta\lb
\phi'(\bmr)-\phi\rb$; $Bg \lb I,\phi\rb$ is the density of spatial
configurations at given $I$ and $\phi$. Given $Bg$ we can compute $W$,
$S$ and its derivatives for {\em any} $\epsilon$ and $\Omega$, e.g.
\begin{eqnarray}
  \nonumber
  \gamma&=&\frac{1}{\beta}\frac{\partial S}{\partial \Omega} \\
  \label{eq:20}
    &=& -\frac{N-5/2}{\beta} \frac{\int dI\:d\phi\: Bg \lb I,\phi\rb I^{-3/2}
  \epsilon_r^{N-7/2}}{W \lb \epsilon,\Omega\rb}.
\end{eqnarray}

The expectation value $\left\langle{\cal O}\right\rangle$ of an observable ${\cal
  O(\bmr)}$ can be estimated if we know $Bg$ and $\left\langle{\cal
    O}\right\rangle_{I,\phi}$
\begin{eqnarray}
  \nonumber
  \left\langle{\cal O}\right\rangle &\equiv& \frac{\int_{v_c}d\bmr\:
    {\cal O}(\bmr) I^{-1/2}
    \epsilon_r^{N-5/2 }}{\int_{v_c}d\bmr\: I^{-1/2} \epsilon_r^{N-5/2}}\\
  \label{eq:21}
  & = & \frac{\int dI \: d\phi \: \left\langle{\cal O}\right\rangle_{I,\phi}
    Bg \lb I,\phi\rb  I^{-1/2} \epsilon_r^{N-5/2}}{W \lb \epsilon,\Omega\rb},  
\end{eqnarray}
where
\begin{equation}
  \label{eq:22}
  \left\langle{\cal O}\right\rangle_{I,\phi} = \frac{\int_vd\bmr\:
    {\cal O}(\bmr)  \delta\lb I'-I\rb
    \delta\lb \phi'-\phi\rb }{Bg \lb I,\phi\rb}.
\end{equation}

Now, we have to compute $Bg(I,\phi)$ and $\left\langle{\cal O}\right\rangle_{I,\phi}$.
{\em A priori\/} $Bg(I,\phi)$ is highly peaked around the values of $I$
and $\phi$ that describe the gas (disordered) phase and should drop down
to the edges. Nevertheless we need a good estimate of $Bg(I,\phi)$ for
almost all values taken by $(I,\phi)$ even when $Bg(I,\phi)$ is very small
comparing to its maximum. For example at small total energy $\epsilon$ only
the part of $Bg(I,\phi)$ for which $\epsilon_r=\epsilon-\frac{\Omega^2}{I}-\phi>0$ will
contribute to the integral~(\ref{eq:19}).

In order to get a good estimate of $Bg$ we used an iterative scheme
inspired by multicanonical
algorithms~\cite{lee93:_new_monte_carlo,berg93:_simul,ferrenberg89:_optim_monte_carlo,smith96:_free_energ_monte_carlo}.
The usual multicanonical task is to compute the free energy as a
function of the total energy. Here we have to compute $Bg$ as a
function of $I$ and $\phi$. The updating scheme presented
in~\cite{smith96:_free_energ_monte_carlo} was used; one of the reason
for this choice is that although it has been given for a one
dimensional task it can be trivially extended to bi--variate problem.
Furthermore we added a blocking mechanism: once we estimate that
enough information has been collected on a given region of the
parametric space $(I, \phi)$ then we tag it as ``locked'' so that it will
not be visited during next iterations.  This mechanism enables the
program to spread more quickly over the parametric space and save
computation time comparing to usual multicanonical algorithms. Details
on this blocking mechanism will be given
elsewhere~\cite{fliegans01:_phase}.

\begin{figure}[h!]
    \epsfig{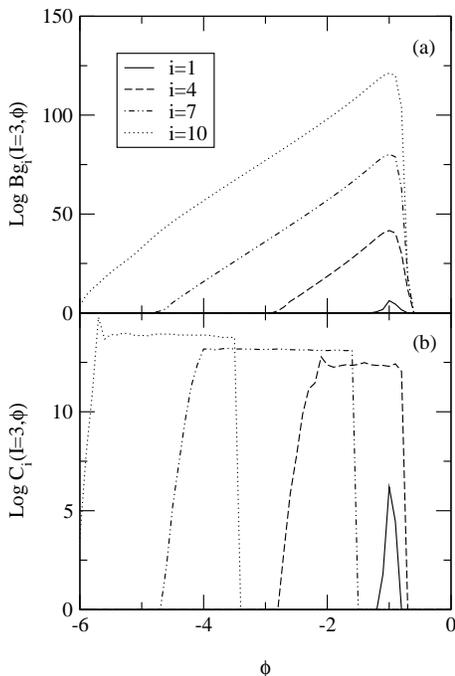}
    \caption{Estimate of the density of state $Bg$ (a) and histogram
      of the visited states $C$ (b) for $I=3$ at different iteration
      steps $i$ of the multicanonical algorithm as a function of the
      potential energy $\phi$. Panel (a) shows how $Bg$ is built step by
      step; $Bg$ is an extremely peaked function, the log of the ratio
      between its maximum and its minimum is about $120$.  Without the
      blocking mechanism (see text) $C_i$ would have been non null for
      all values of $\phi$ visited during previous steps $j<i$. In panel
      (b) we see that the algorithm does no longer visit
      ``well--known'' regions ($\phi\gtrsim-1.5$) already after four steps.}
    \label{fig:muca_hst_bg}
\end{figure}

In the present paper we present results for $N=20$, and $\sigma=0.05$.
Figure~\ref{fig:muca_hst_bg} shows a slice of $Bg(I,\phi)$ for $I=3$ at
different iteration steps $i$ (Fig.~\ref{fig:muca_hst_bg}.a). We have
also plotted the histogram $C(I=3,\phi)$ of the visited region in order
to illustrate the blocking mechanism (Fig.~\ref{fig:muca_hst_bg}.b).
As expected $Bg$ is strongly peaked around the disordered region
$\phi\approx-1$ (this value correspond to the mean of $\phi$ over randomly
generated spatial configurations); after 10 iterations the ratio
between the maximum and the minimum of $Bg$ is $\approx\exp 120$. This ratio
increases exponentially with $N$, e.g. at $N=10$ its value is $\approx \exp
80$. This is the main reason why we could not study systems with
larger $N$ within the current algorithm, but we should add that no
qualitative changes have been noted between $N=10$ and $N=20$ (some
preliminary studies support this remark for $N=50$).

\section{Results}
\label{sec:results}

\subsection{Entropy and its derivatives}
\label{sec:entropy-other-things}

\begin{figure}[h!]
  \begin{center}
    \epsfig{file=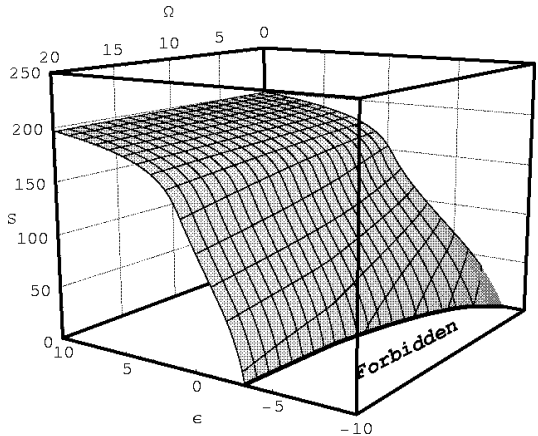,width=70mm}
  \end{center}
  \caption{Entropy surface $S\lb \epsilon,\Omega\rb$, the mesh lines are at
    constant $\epsilon$ or constant $\Omega$. The thick line is close to the
    pojection of the $T=0$ ($S\to-\infty$) isotherm. A convex intruder at
    constant $\Omega$ and for a certain energy range (e.g. $-2<\epsilon<0$ for
    $\Omega=20$) can be seen for all $\Omega$. $S$ is not defined in the {\em
      forbidden region\/}; there the remaining energy $\epsilon_r$ is
    negative for all ${\bf q} \in v_c $.}
  \label{fig:SSurf}
\end{figure}

Figure~\ref{fig:SSurf} shows the entropy surface $S$ as a function of
$\epsilon$ and $\Omega$. The ground state energy $\epsilon_g(\Omega)$ (thick line in
Fig.~\ref{fig:SSurf}) increases with $\Omega$; $\epsilon_g$ classically corresponds
to $\epsilon_r=0$ implying $S=-\infty$. For all $\Omega$, $\epsilon_g(\Omega)$ is a concave
function of $\Omega$, i.e. $\frac{\partial^2\epsilon_g}{\partial\Omega^2} \leq 0$; at high $\Omega$ ($\Omega\gtrsim12$)
it is almost linear $\frac{\partial^2\epsilon_g}{\partial\Omega^2} \to 0^-$. These properties show
that the set $\{\epsilon,\Omega\}$ over which $S$ is defined is {\em not\/}
convex~\cite{convex}, and in Sec.~\ref{sec:beta-beta-gamma} we discuss
some consequences resulting from a non--convex domain of definition.

\begin{figure}[h!]
  \begin{center}
    \epsfig{file=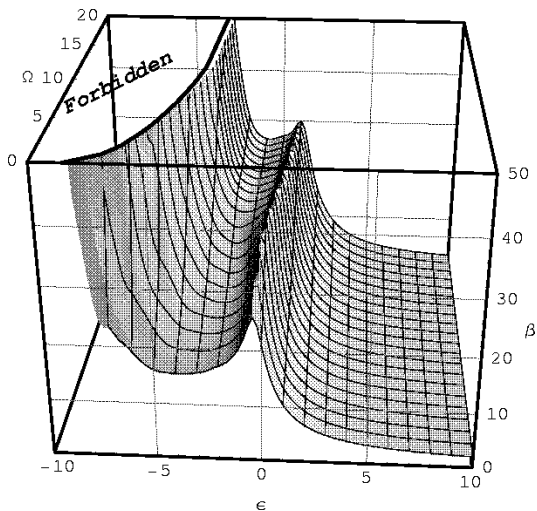, width=70mm}
  \end{center}
  \caption{Inverse temperature $\beta(\epsilon,\Omega)$ surface. The mesh lines are at
    constant $\epsilon$ or constant $\Omega$. The intruder in $S$ at fixed $\Omega$
    corresponds here to a multiple energy value for a given $\beta$ and
    $\Omega$, e.g. $\beta(\epsilon,\Omega=0)=20$ has three solutions $\epsilon_1\approx0$, $\epsilon_2\approx-1$ and
    $\epsilon_3\approx-6$. The thick line is close to the projection of the $\beta=\infty$
    isotherm; $\beta$ is not defined in the {\em forbidden\/} region.}
  \label{fig:BetaSurf} 
\end{figure}

At fixed $\Omega$, $S(\epsilon)$ is not concave for all $\epsilon$ but
shows for some energy interval a convex intruder which signals a first
order phase 
transition with negative specific heat
$\lb \frac{\partial\beta}{\partial\epsilon}<0 \rb$\cite{gross97:_microc}.  This can be better viewed by
plotting $\beta(\epsilon,\Omega)=\frac{\partial S}{\partial \epsilon}$
(Fig.~\ref{fig:BetaSurf}). Here the counter part of  
the entropy--intruder is a region of multiple valued $\epsilon(\beta)$, this is
the case for $\beta$ between 15 and 20.
The latent heat at fixed $\Omega$, $q_\epsilon(\Omega)$ decreases for
$0\leq\Omega\lesssim12$ and is a 
constant for $\Omega>12$. There is no critical value of $\Omega$,
$\Omega_c$ above which $S(\epsilon)$ is concave for all $\epsilon$,
i.e. there is a first order phase transition of all values of
$\Omega$. In another model for a self--gravitating system such a $\Omega_c$ was
reported~\cite{laliena99:_effec}, but not in the one presented
in~\cite{klinko00:_mean_field_theor_spher_gravit_system}. 

\begin{figure}[h!]
  \begin{center}
    \epsfig{file=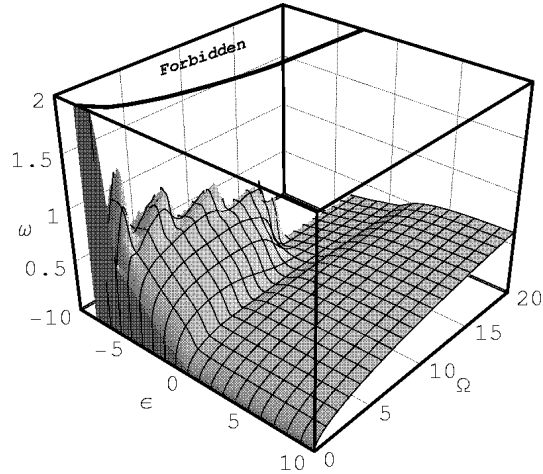, width=70mm}
  \end{center}
  \caption{Microcanonical angular velocity 
    $\omega\lb \epsilon,\Omega \rb$ surface. The mesh lines are at constant $\epsilon$ or
    constant $\Omega$. $\omega$ is not defined in the {\em forbidden\/} region.
    At high energy $\omega\propto\sqrt{\Omega}$; Near the ground states $\omega$ shows a
    richer non--monotonic behavior with peaks and troughs for small
    $\Omega$ and has a nearly constant value for large $\Omega$ (see text).}
  \label{fig:WSurf}
\end{figure}

On Fig.~\ref{fig:WSurf} we have plotted the microcanonical angular
velocity $\omega$ as a function of $\Omega$ and $\epsilon$. As a direct consequence of
Eq.~(\ref{eq:8}) $\omega$ tends to zero with $\Omega$, and at high energy $\omega$ is
proportional to $\sqrt{\Omega}$ ($\propto L$). For low energies and
$\Omega<12$, $\omega$ exhibits some structures with peaks and troughs, in
another words at fixed $\epsilon$, $\omega$ is not necessarily an increasing
function of $\Omega$. At high $\Omega$ ($\Omega>12$) and near the ground states $\omega$
is almost a constant. All these structures can be understood in terms
of mass distributions which influence $\omega$ through $I$ (see
Sec.~\ref{sec:mass-distribution}).

\subsection{Mass distribution}
\label{sec:mass-distribution}

In order to understand the origin of the structures seen in the
different microcanonical quantities ($S$, $\beta$, $\omega$, $\ldots$) we have to
have a closer look at the spatial configurations, i.e. at the mass
distributions.
One observable we have studied is the mass density $\rho$ (see
Eq.~\ref{eq:21} and~\ref{eq:22}). As the Hamiltonian of the system is
rotationally invariant, the mean value of $\rho$ can only be a function
of $r$, the distance from the center of coordinates, although other
observables might show a breaking of the rotational symmetry (see below).

\begin{figure}[t!]
  \begin{center}
    \epsfig{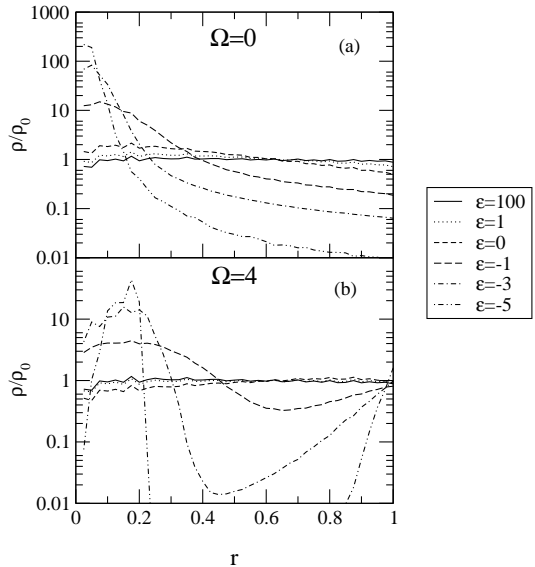}
  \end{center}
  \caption{Density as a function of the distance from the center $r$
    at different 
    values of energy $\epsilon$ and angular velocity $\Omega$ (arbitrary
    units). At high energy and 
    for all $\Omega$ the density is flat; the system is in the homogeneous
    gas phase. Near the ground state the density shows one peak for
    $\Omega=0$ (a) and two peaks for $\Omega>0$ (b), which correspond respectively to a one
    cluster and to a two clusters phase surrounded by some gas (see text).}
  \label{fig:rho1}
\end{figure}

On Fig.~\ref{fig:rho1} $\rho$ is plotted for different energies and for
$\Omega=0$ and $\Omega=4$. For $\Omega=0$~(Fig.~\ref{fig:rho1}.a) we recover the
classical case (when $E$ is the only fixed parameter); at high energy
the system is in a homogeneous gas phase (flat $\rho$), when the energy
decreases the system undergoes a phase transition and eventually ends
up in a collapse phase where a majority of particles are in a cluster
near the center of coordinates ($\rho$ peaked at $r=0$).  For
$\Omega\neq0$~(Fig.~\ref{fig:rho1}.b) the situation is (a little bit)
different.  At high energy we recover the homogeneous gas phase. But
at low energy the system cannot collapse entirely at the center of
mass. This is due to the rotational energy $\epsilon_{rot}=\frac{\Omega}{I}$ in
Eq.~(\ref{eq:5}); if the system contracts at the center the inertial
momentum $I$ will tend to zero and therefore $\epsilon_{rot}$ will diverge
leading to a negative remaining energy $\epsilon_r$. So depending on the
value of $\Omega$ the main cluster will eject a certain amount of particles
in order to increase $I$. Near the ground state these ``free''
particles will eventually collapse to form a second cluster in order
to decrease the potential energy $\phi$. Due to the conservation of the
center of mass, the position of the biggest cluster will be shifted
from the center by a certain amount (see Fig.~\ref{fig:rho1}.b at
$\epsilon=-5$).  At low $\Omega$ one particle will be ejected, with increasing $\Omega$
the number of ejected particles raises and this process stops when two
equal--size clusters are formed. This explains the discreteness of the
peaks in $\omega$ (Fig.~\ref{fig:WSurf}), the increase of the ground state
energy $\epsilon_g(\Omega)$; because the potential energy of a single cluster of
size $N$ is smaller than the one of two well separated clusters. At
high $\Omega\gtrsim12$ the system undergoes a phase transition from a gas phase
to a collapse phase with two equal size clusters close to the
boundary. From one value of $\Omega=\Omega_1>12$ to another one $\Omega_2>\Omega_1$ the
whole entropy curve at fixed angular momentum is simply shifted along
the energy axis, i.e. $S\lb \epsilon, \Omega_1 \rb \approx S\lb \epsilon + \frac{\Omega_2-\Omega_1}{N},
\Omega_2 \rb $.  So the ground state energy $\epsilon_g(\Omega)$ at high $\Omega$ is almost
on a line of equation $\epsilon_g+\frac{\Omega}{N}+\phi_g\approx0$, where $\frac{\Omega}{N}$ and
$\phi_g$ are the rotational energy and the potential energy of 2 clusters
of size $N/2$ at radius $r=1$ respectively.  This monotonic behavior
has already be mentioned for all the thermodynamical variables $S$
(Fig.~\ref{fig:SSurf}), $\beta$
(Fig.~\ref{fig:BetaSurf}), $\omega$ (Fig.~\ref{fig:WSurf})\ldots

As already mentioned $\rho$ is only a function of $r$ and it can not be
used to infer the angular distribution of the particles, i.e. there is
not enough information to say if a peak in $\rho$ at $r_0\neq0$ corresponds
to one or many clusters or to a uniform distribution of the particles
(ring) lying on a circle of radius $r_0$. However at least at very low
energy a many clusters (more than two) configuration is very unlikely
and will not contribute to the average values for reasons linked to
the weight $W(\bmr)=\frac{1}{\sqrt{I}}\epsilon_r^{N-5/2}$. For simplicity let
us assume that there is only one strong peak in $\rho$ at $r=r_0\neq0$.
Since the center of mass is fixed this can not be the signature of a
1--cluster system. At least 2--clusters lying on a circle of radius
$r_0$ are needed. All these n--clusters systems have the same
rotational energy $\frac{\Omega}{I}=\frac{\Omega}{Nr_0^2}$, but their
corresponding potential energy $\phi_n$ will differ. For example, with
$\sigma=0.05$, $r_0=0.5$ and $N=24$ the ratio of potential energy is
$\frac{\phi_2}{\phi_3}\simeq1.7$. So at low energy, the remaining energy $\epsilon_r$
corresponding to a 2--clusters system will be much larger than the
3--clusters' one, leading to a {\em huge\/} difference in the weight
$W(\bmr)$. So at low energy and for $\Omega\neq0$ the 2--clusters case is
dominant. At higher energies, the term $Bg(I,\phi)$ in Eq.~(\ref{eq:19})
can compensate the difference in the weight $W(\bmr)$ and allow many
clusters configurations and eventually at high energy a complete random
configuration on the ring of radius $r_0$ will dominate the average
mass distribution. 

\begin{figure}[h!]
  \begin{center}
    \epsfig{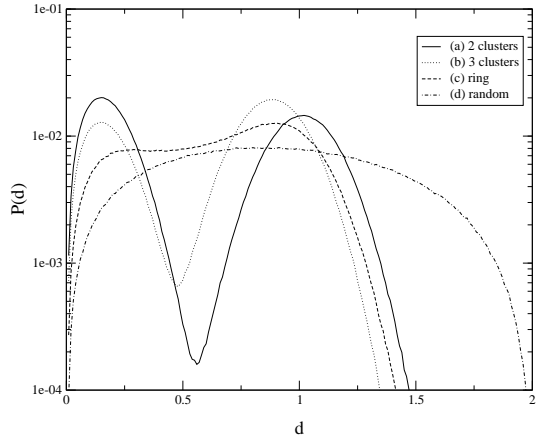}
  \end{center}
  \caption{Average of $P(d)$ the distance distribution for different
    simulated spatial configurations. See text.}
  \label{fig:checkdist}
\end{figure}

We can check this argument by studying another observable, for example
the normalized distance distribution $P(d)$, i.e. the density of
probability that the distance between two given particles is $d$. To
probe the information given by $P(d)$ we have estimated it for four
simple mass distributions: (a) 2--clusters, (b) 3--clusters, (c) ring,
(d) uniform random distribution. For (a), (b) and (c) the particle
were put on a circle of radius $r_0=0.5$, and then randomly shifted
several times (in order to give a spatial extension to these idealized
initial configurations); finally the $\frac{N(N-1)}{2}$ distances are
recorded for all realizations and averaged. Figure~\ref{fig:checkdist}
shows the average of $P(d)$ over 1000 realizations. Note that the
density distribution $\rho(r)$ is by construction exactly the same for
the three first cases, i.e. strongly peaked at $r_0$ with a width of
about $0.5$. The latter value depends on the shift one applies on the
initial idealized spatial configurations.

As one can see on Fig.~\ref{fig:checkdist}, although the density
distribution is the same for (a), (b) and (c), $P(d)$ gives some new
insight on the mass distribution:
\begin{itemize}
\item[(a)] There are two peaks, one at small $d$ which corresponds to
  a clusterisation and another one at $r=1=2r_0$; this is exactly the
  distance between the two clusters (more precisely between their
  center of mass). The area under the small $d$ peak is similar to the
  one under the large $d$ peak, indeed the number of short distance
  pairs and the number of pairs with $d\simeq1$ are both about
  $\frac{N^2}{4}$. Moreover the widths of the first and second peaks
  are (as expected) $\sim0.5$ and $\sim1=2*0.5$ respectively.
\item[(b)] There are again two peaks one at small $d$ and another at
  $d\simeq0.8<1$ and their respective widths are similar to the ones in
  (a). The large $d$ is compatible with the length of one side of the
  equilateral triangle on top of which the 3--clusters mass
  distribution has been built. This time the area under the large $d$
  peak is larger than the one under the short $d$ peak, since the
  number of short distance pairs is about $\frac{N^2}{6}$ whereas the
  number of pairs with $d\simeq0.8$ is $\frac{N^2}{3}$.
\item[(c)] In the ring a trace of the two peaks still exists but they
  are not well separated because a lot of intermediate distances are
  compatible with this model.
\item[(d)] When the particles are uniformly distributed $P(d)$ is a
  binomial--like distribution.
\end{itemize}

\begin{figure}[h!]
  \begin{center}
    \epsfig{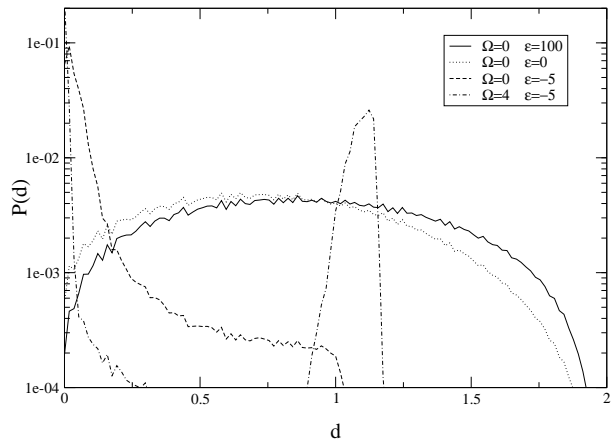}
  \end{center}
  \caption{Distance distribution $P(d)$ for the gravitational case for
    different values of $(\epsilon,\Omega)$. At high energy $P(d)$ corresponds to
    a random distribution (see Fig.~\ref{fig:checkdist}). For $\Omega=0$
    and at low energy, $P(d)$ has one peak at $d\approx0$; almost all
    particles are very close from each other, and there is a single
    cluster collapse phase. For $\Omega\neq0$ there are two peaks at low
    energy: one at very small $d$ which is a sign of clusterisation
    and another peak at large $d$ which signals multiple
    clusterisations; in fact there are two clusters (see text).}
  \label{fig:dist}
\end{figure}

We have also estimated $P(d)$ for our gravitational system, as shown
on Fig~\ref{fig:dist}. At high energy, $P(d)$ corresponds to the
randomly distributed case (see Fig.~\ref{fig:checkdist}). At low
energy with $\Omega=0$, $P(d)$ has only one peak at $d=0$, this corresponds
clearly to a single cluster case surrounded by some gas. For $\Omega\neq0$ and
at low energy (in Fig.~\ref{fig:dist} $\epsilon=-5$ and $\Omega=4$), there are two
well separated peaks, one at small $d_0=0$ and the other at $d_1\simeq1.1$.
The peaks imply the presence of at least two clusters, however the
fact that the widths of the peaks are small excludes a large number of
clusters and even more the ring case (see Fig.~\ref{fig:checkdist}).
Now we can combine these informations with the ones obtained by
studying $\rho(r)$ (see Fig.~\ref{fig:rho1}). For $\epsilon=-5$ and $\Omega=4$, $\rho$
has two peaks at $r_1\simeq0.15$ and $r_2\simeq1$. All in all, this means that
there are, in the mean, two clusters rotating around the center of
mass. The distance between these clusters is $r_1+r_2\simeq1.15\simeq d_1$.
Their mass ratio is $\frac{m_2}{m_1}=\frac{r_1}{r_2}\simeq0.15$. Since we
know the total mass $m_1+m_2=20$ we get $m_1\simeq17$ and $m_2\simeq3$.

The distance distribution can be of great help to identify the mass
distributions at low energies. However at the transition regions since
there is a superposition of different types of mass distributions the
knowledge $\rho$ and $P(d)$ is not sufficient and therefore of no help if
we want to study for example the ``fractality'' of the mass
distribution as it has been discussed in other self--gravitating
systems
~\cite{vega00,sota00:_origin_fract_distr_self_gravit,semelin99:_physiq}
(not mentioning the fact that the actual number of particles in the
presented numerical applications is too  small), and further work is
needed to get a more detailed picture.

At very low energy, near the ground state at least one of the clusters
(the smallest) is very close to the boundary. There the assumption of
a small evaporation rate made in Sec.~\ref{sec:introduction} does not
hold.

\subsection{Phase diagram}
\label{sec:phase-diagram}

\begin{figure}[h!]
  \begin{center}
    \epsfig{file=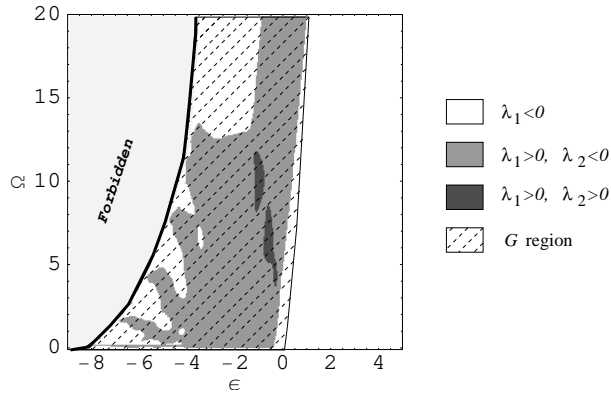, width=80mm}
  \end{center}
  \caption{$\sign \lb \lambda_1(\epsilon,\Omega) \rb$ as defined in
    Sec.~\ref{sec:phase-phase-trans}. The white regions correspond to
    $\lambda_1<0$ and therefore to a positive determinant $D_S=\lambda_1\lambda_2>0$ of
    the entropy curvature, these are pure phase regions.  The gray
    region corresponds to $\lambda_1>0$ and $D_S<0$ and the dark gray ones
    also to $\lambda_1>0$ but $D_S>0$; $\lambda_1>0$ defines first order phase
    transition regions (see text). Points in $G$ correspond to local
    maxima (minima) in Eq.~(\ref{eq:37}) if $\lambda_1<0$ ($\lambda_1>0$) but not
    to global maxima (minima); points outside $G$ correspond to global
    maxima in Eq.~(\ref{eq:37}).  There is a one--to--one mapping
    between the microcanonical ensemble and the GBE only in the white
    region (see text). $S$ is not defined in the {\em forbidden\/}
    region, here in light gray.  Note that (a) the points at $\Omega=0$ and
    low energies $\epsilon < -7$ are not included in $G$, (b) the high energy
    limit of $G$ is only known approximatively.
    }
  \label{fig:sdet}
\end{figure}

On Fig.~\ref{fig:sdet} we have plotted $\sign \lb \lambda_1(\epsilon,\Omega) \rb$ as
defined in Sec.~\ref{sec:phase-phase-trans}. This plot can be taken as
the phase diagram of the self--gravitating system at fixed $\epsilon$ {\em
  and\/} $\Omega$. The white regions correspond to pure phases ($\lambda_1<0$ ). At
high energy there is a homogeneous gas phase and at low energy there
are several pure collapse phases with one ($\Omega=0$) or two ($\Omega\neq0$)
clusters.  The different 2--clusters phases are characterized by the
relative size of their clusters (see
Sec.~\ref{sec:mass-distribution}). These regions are separated by
first order phase transition regions where $\lambda_1>0$ (gray in
Fig.~\ref{fig:sdet}). There are even two regions (dark gray) where the
entropy $S$ is a convex function of $\epsilon$ and $\omega$; i.e. all the
eigenvalues of $J_S$ are positive ($\lambda_1>0$ {\em and} $D_S>0$). These two
regions are quite stable with respect to the number of particles (at
least for $N<50$). Their surface increase slightly with the number of
particles $N$.  The orientation of ${\bf v}_1$ the eigenvector
associated with $\lambda_1$ (defined as the largest eigenvalue of $J_S$) is
not yet known in detail for all $(\epsilon,\Omega)$; however we can already state
that at ``high'' energy ${\bf v}_1$ is almost parallel to the energy
axes (phase transition in the $\epsilon$ direction) and should be parallel to
the ground state at very low energy. The overall structure of the
collapse phases matches the one of the angular velocity $\omega$ (see
Fig.~\ref{fig:WSurf}): roughly, the peaks in $\omega$ correspond to pure
phases while the valleys between these peaks belong to the first order
phase transition region. Due to a lack of precision we are yet unable
to ensure that the two isolated pure phases at $\epsilon\approx-4$, $\Omega\approx1$ and $\Omega\approx7$
are really isolated and not linked to a pure phase at lower energies. 

\begin{figure}[h!]
  \begin{center}
    \epsfig{file=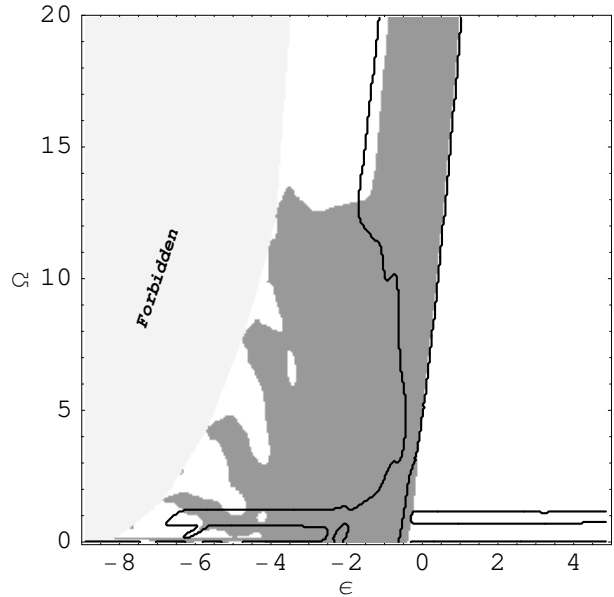, width=80mm}
    
    \caption{Locus of second order phase transitions (see text).}
    \label{fig:second-order}
  \end{center}
\end{figure}
As already mentioned, unlike in the model presented by Laliena in
\cite{laliena99:_effec} there is no critical angular momentum $L_c$
above which the first order phase transition vanishes giving rise to a
second order phase transition at $L_c$. Nevertheless this does not
exclude second order phase transition (critical point) at all. They
are defined in the microcanonical ensemble by: (i) $\lambda_1=0$; (ii)
$\nabla\lambda_1\cdot{\bf v}_1=0$ (see Sec~\ref{sec:phase-phase-trans}). On
Fig.~\ref{fig:second-order} (just like on Fig.~\ref{fig:sdet}) regions
where $\lambda_1<0$ ($>0$) are in white (gray). The condition (i) is simply
achieved at the boarder between the gray and the white regions. The
thick lines on Fig.~\ref{fig:second-order} correspond to condition
(ii). Second order phase transition are located at the crossing points
points of the thick lines and the boarders. One immediately sees that
there are several critical points. However there are not all of
(astro--)physical interest since most of them are close to the ground
states line or at very high angular momentum where the small
evaporation rate assumption is not valid. Nevertheless there are two
points one at $(\epsilon,\Omega)\approx(-0.5,1)$ and another one at $(\epsilon,\Omega)\approx(-0,4)$ where
this assumption is valid and therefore their deserve further
investigations and especially regarding their corresponding mass
distributions.

\subsection{Loss of information in the $(\beta, \beta \gamma)$ ensemble (GBE)}
\label{sec:beta-beta-gamma}

In a recent paper Klinko and Miller have studied another model for
rotating self--gravitating
systems~\cite{klinko00:_mean_field_theor_spher_gravit_system}. They
introduced the canonical analogous of the $(\epsilon,\Omega)$ ensemble namely the
$(\beta,\gamma\beta)$ ensemble (GBE), see Eq.~(\ref{eq:6}) and (\ref{eq:9}). Once
one defines the generalized microcanonical ensemble (ME), it is
straightforward to introduce and study the system in its conjugate
ensemble, the generalized canonical ensemble (CE).  Note that
``generalized'' means in CE ``as a function of all the intensive
variables'' and, in ME ``as a function of all extensive parameters''.
There are mainly two reasons invoked to study a system in CE instead
of in ME
\begin{enumerate}
\item[1.] Performing the computations in CE are in most cases much
  easier than in ME. Here CE can be seen as a trick
  \cite{ehrenfest12a}. However there is a priori no reason for the
  results to be equivalent. Indeed the strict equivalence of the
  ensemble is only achieved at the thermodynamical limit except in the
  first order phase transition
  regions~\cite{gross97:_microc,gross00:_phase_trans_small,landau94:_physiq}.
  In a weaker sense, needed for Small systems, ME and CE can only be
  equivalent when $\lambda_1(X)<0$ and far from any first order phase
  transition regions (see below). Moreover recent progress in computer
  performance enables one to perform now numerical experiments within
  the ME for increasingly complex systems.
\item[2.] The studied system is not isolated but is in contact with a
  ``heat bath'' and exchange amounts of $X$ with it. Then, obviously
  ME does not provide a suitable description, and CE might be
  eligible. Here CE describes a {\em different\/} physical system than
  the one in ME. At the thermodynamical limit (if it exists) the
  ensembles are again equivalent except at first order phase
  transitions. However, the CE description is valid only if the
  Hamiltonian of interaction ${\cal H}_{int}$ is small comparing to
  the ones of the system ${\cal H}_{sys}$ and the heat bath ${\cal
    H}_{hb}$. This condition is usually fulfilled at the
  thermodynamical limit (if it exists), but for Small ones ${\cal
    H}_{int}$ can hardly be small at least comparing to ${\cal
    H}_{sys}$ and this can lead to dramatic
  effects~\cite{sato00:_irrev,sekimoto00:_carnot_cycle_small_system}.
  In this case a better description would be the ME of the system, its
  ``heat bath'' {\em and\/} their interactions.
\end{enumerate}

For now on we consider in this paper the cases when CE is used as a
mathematical trick, and we focus and the amount of information lost
from ME to CE.

The link between ME and CE is given by Laplace transform (using the
notations introduced in Sec.~\ref{sec:phase-phase-trans})
\begin{equation}
  \label{eq:37}
  {\cal Z}(x)=\int _0^{\infty}dX \: e^{-X\cdot x+S(X)},
\end{equation}
where ${\cal Z}$ is the partition sum of the CE; $x=\{x^{1},\ldots
,x^{{\cal N}}\}$ are the intensive variables associated with $X$ and
defined by
\begin{equation}
  \label{eq:25}
  x^{i}\equiv\frac{\partial S}{\partial X^i},
\end{equation}
for $i=1,\ldots,{\cal N}$. The ME at $X_0$ is equivalent to the CE at
$x_0=x(X_0)$ if the integrand in Eq.~(\ref{eq:37}) $e^{f(X,x_0)}$, and
therefore $f(X,x_0)=-X\cdot x_0+S(X)$ has a {\em global maximum\/} at
$X_0$. This condition is violated when $\lambda_1(X_0)>0$ (this is the basic
idea behind the definition of phase transitions in Small systems,
see~\cite{gross00:_phase_trans_small}). So in practice all
informations contained in points where $\lambda_1>0$ are lost after the
Laplace transform~(\ref{eq:37}) in CE.

\begin{figure}[h!]
  \begin{center}
    \epsfig{file=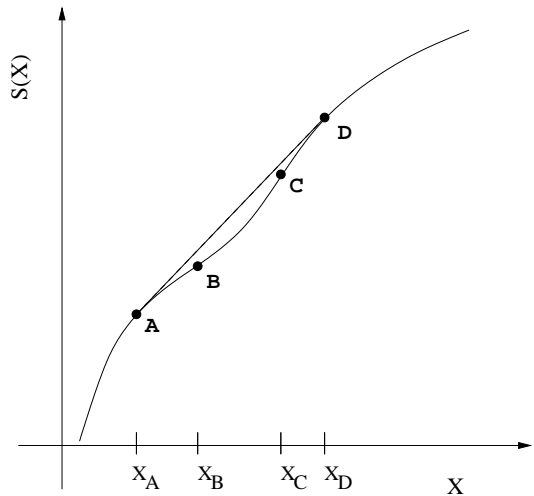, width=70mm}
  \end{center}
    \caption{Schematic entropy curve $S$ as a function of one extensive
      parameter $X$ (solid line) and the Maxwell line (dashed line).
      $A$ and $D$ are the end points of the Maxwell construction. At
      $B$ and $C$ the largest curvature vanishes, i.e. $\lambda_1=0$.  All
      the points below the Maxwell line do not not correspond to a
      global maximum in the Laplace transform Eq.~(\ref{eq:37}). This
      includes points where $\lambda_1=\frac{\partial^2S}{\partial X^2}>0$, but also
      points where the entropy is concave, for $X\in[X_A,X_B]$ and
      $X\in[X_A,X_B]$, $\lambda_1(X)<0$.}
  \label{fig:SEx}
\end{figure}

Now we are left with the points $X_0$ characterized by $\lambda_1(X_0)<0$;
this relation implies only that $f(X,x_0)$ has a local maximum at
$X_0$ but not that it is a {\em global\/} one; $\lambda_1<0$ is a necessary
but not sufficient condition. In Fig.~\ref{fig:SEx} we have
illustrated this point with a trivial one--dimensional example; all
the points below the Maxwell line do not correspond to global maxima
and their information content is smeared out and, in practice, lost in
CE.

From the last remark we see that a way to check if a point corresponds
to a global maximum in Eq.~(\ref{eq:37}) is to study what would be a
generalization of the Maxwell line in ${\cal N}>1$ dimensions. Work is
in progress in this direction and the results will be presented
elsewhere, but we can already state that this task is, to some extent,
similar to the one of building the {\em convex hull\/} of a set of
points in ${\cal N}$ dimensions, or more specifically the convex hull
of the entropy $S$.

However, if one needs qualitative results, in two dimensions the task
can be rather easily solved in another way. Let us define $p(X,X_0)$
the tangent plane to $S(X)$ at $X_0$; its equation is $p(X,X_0)=-X\cdot
x_0+S(X_0)+X_0\cdot x_0$. If $f(X,x_0)$ has a global maximum at
$X_0$ then
\begin{equation}
  \label{eq:38}
  p(X,X_0)>S(X) 
\end{equation}
for all $X\neq X_0$ and $p(X_0,X_0)=S(X_0)$ by definition.

In the case of the gravitational model presented here $X=\{\epsilon,\Omega\}$ and
$x=\{\beta,\gamma\beta\}$.  Now if one inspects the entropy surface $S(X)$ (see
Fig.~\ref{fig:SSurf}) it is clear that condition (\ref{eq:38}) is not
satisfied for all the points in the region filled with dashed lines
($G$) in Fig.~\ref{fig:sdet}. This is due to the concavity of the
energy ground state $\epsilon_g(\Omega)$ (see
Sec.~\ref{sec:entropy-other-things}). $G$ includes all the
two--clusters collapsed phases, the first and second order phase
transitions (except for $\omega=\Omega=\gamma=0$). All the information contained in
$G$ is smeared out through the Laplace transform (\ref{eq:37}) in GBE
and, in practice, lost.

\begin{figure}[h!]
  \begin{center}
    \epsfig{file=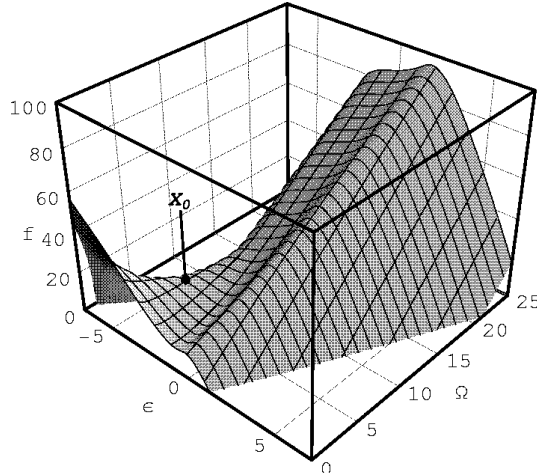, width=70mm}
    
    \caption{$f(\epsilon,\Omega,\beta_0,\gamma_0) = -\epsilon\beta_0  -\Omega\beta_0\gamma_0 +S(\epsilon,\Omega) - K$
      as a function of $\epsilon$ and $\Omega$, where $K$ is an arbitrary
      constant; $\beta_0= \beta(\epsilon_0,\Omega_0)$; $\Omega_0= \Omega(\epsilon_0,\Omega_0)$; $X_0= (\epsilon_0,\Omega_0)=
      (-3,5) \Rightarrow (\Omega_0,\gamma_0)\approx (27.9,-0.196)$. The mesh lines are at
      constant $\epsilon$ or constant $\Omega$. As expected $f$ has a saddle point
      at $X_0\approx (\beta_0,\gamma_0)$ since $m(X_0)>0$ and $D_S(X_0)<0$ (see text
      and Fig.~\ref{fig:sdet}). $f$ has a global maximum at $\Omega=0$ and
      $\epsilon\approx-7$, but one sees that it is an monotonically increasing
      function for increasing $\Omega$ and $\epsilon(\Omega)\approx \epsilon_g(\Omega)+2$. Therefore the
      integral in Eq.~(\ref{eq:37}) {\em diverges\/} and the $(\beta,\gamma\beta)$
      ensemble is {\em not defined\/} for $(\beta_0,\gamma_0)$.}
    \label{fig:N20.f}
  \end{center}
\end{figure}

The fact that GBE misses all the two--clusters collapse phases would
be already enough to disqualify it as being a good approximation
(mathematical trick) of the ME. But, furthermore, if one studies more
carefully $f(X,x_0)$, $X_0 \in G$; one will notice that (a) there is one
local maximum at $\Omega=0$ and (b) there is no maximum for high $\Omega$: in
the direction of increasing $\Omega$ at low energy, $f(X,x_0)$ is a never
ending increasing function, i.e. $f(X,x_0)$ has {\em no global
  maximum\/} for $X\in G$ (see Fig.~\ref{fig:N20.f}). Therefore the
integral in Eq.~(\ref{eq:37}) {\em diverges\/} for all $x_0$, $X_0\in
G$. In other words the GBE, in our model, is {\em not defined\/} for
high $\beta$ and $\gamma\neq0$ ($\omega\neq0$).

This example shows how dramatic can be the information loss if one
studies an isolated system in the CE.

\section{Summary and discussion}
\label{sec:comments-conclusions}

\subsection{Summary}
\label{sec:summary}

The aim of this paper is to present the results of a study of a
self--gravitating system of $N$ classical particles on a disk of
radius $R$ for which two extensive variables, namely total energy $E$
and total angular momentum $L$ are fixed. The microcanonical entropy
$S$ can be written as the logarithm of an integral over the spatial
configurations (\ref{eq:3}) and (\ref{eq:5}). The conservation of $L$
implies that the mean value of the linear momentum $\vp$ of a particle
at a radial distance $r$ is proportional to $L$ and $r$ (\ref{eq:11});
in the mean the system rotates like a solid body, i.e. the mean
angular velocity of a particle does not depend on its position
(\ref{eq:15}). The dispersion of $\vp$ is broader than the one it had
if $L$ would not be conserved and depends on the radial position
(\ref{eq:16}). In order to integrate $e^S$ we write it as the folding
product of a ``background'' function $Bg$ and the weight associated
with the remaining energy (\ref{eq:19}).  Once a numerical estimate of
$Bg$ is obtained, $S$ and its derivatives, such as the inverse
temperature $\beta$, can be computed. The entropy surface shows an
intruder in the energy direction which signals a first order phase
transition with negative specific heat (Fig.~\ref{fig:SSurf} and
Fig.~\ref{fig:BetaSurf}). Contrary to another
model~\cite{laliena99:_effec} there is no critical value of $L$ above
which this transition is no longer present, nevertheless there are
several points of the parametric space $\lb E, L\rb$ . At high energy
the angular velocity $\omega$ is a simple increasing function of $L$
(Eq.~(\ref{eq:8}) and Fig.~\ref{fig:WSurf}), but at low energy, near
the ground states the relation between $L$ and $\omega$ becomes
non--trivial. All these peculiarities can be understood if we study
the mass distribution. We use two observables: the density of mass $\rho$
as a function of the radial distance; and $P(d)$ the density
probability that the distance between two particles is $d$. At fixed
$L$ and at high energy the particles are randomly distributed over the
disk (gas phase); if the energy decreases, at some threshold the
system undergoes a phase transition of first order to a collapse
phase. If $L=0$ there is a single cluster at the center of mass, if
$L\neq0$ then there are two clusters rotating around the center of mass.
Their relative mass $\frac{m_1}{m_2}$ with $m_1>m_2$ is very large for
low $L$, it decreases as $L$ raises and eventually becomes 1 for $L$
larger than a certain threshold. Clearly these two clusters phases
were not reported in previous work because of the usual {\em a
  priori\/} assumption of rotaional symmetry.

Phase transitions (and phases) can be defined unambiguously in the
microcanonical ensemble as a function of the {\em local\/} topology of
$S$. If all eigenvalues of $J_S$, the Jacobian of $S$ are negative
then the system is in a pure phase. If (at least) one eigenvalue is
positive, say $\lambda_1$, then there is a first order phase transition in
the direction of ${\bf v}_1$, the eigenvector associated with $\lambda_1$. A
critical behavior occurs when one (or many) eigenvalue vanishes on at
least a second order region (if $\lambda_i=0$, $\nabla_{{\bf v}_i}\lambda_i=0$). Note
that one can find positive, negative and zero eigenvalues at the same
point in the parameter space. Using these criteria, we can draw a
phase diagram of the gravitational system (Fig.~\ref{fig:sdet}) at
fixed $E$ {\em and \/} $L$: at high energy there is a pure gas phase;
at low energy there are several pure collapse phases, there is one
cluster at $L=0$ and there are two clusters for $L\neq 0$; these phases
are separated by first order phase transition regions. There are also
several second order phase transitions; two of them are located at
relativley high energies ($E\approx0$) and therefore may be of astrophysical
importance.

Studying an isolated system using the canonical ensemble (CE) can be
very misleading, since there is a massive loss of information from the
correct microcanonical ensemble (ME) description to the CE's one (if
there are phase transitions in ME). In fact for the gravitational
model, {\em CE cannot be sensitive to the two clusters phases and all the
  phase transitions\/} and {\em it is not defined for some values of
  the intensive parameters that exist in ME\/} (see
Sec.~\ref{sec:beta-beta-gamma}).

\subsection{Discussion}
\label{sec:discussion}

Of course we have just presented an {\em equilibirum statistical\/}
model that may help to understand the physics of globular clusters or
collapsing molecular clouds and the results should be interpreted with
caution especial in the case of star formation. A lot of
``ingredients'' are missing in order to have a complete picture of the
formation of multiple stars systems and planetary systems, for instance the magnetic
field~\cite{2000NewA....4..601H,galli00:_singul_isoth_disks_format_multip_stars},
or the presence of vortices~\cite{2000A&A...356.1089C}.

Phases and phase transitions can be well defined in the microcanonical
ensemble (ME) without invoking the thermodynamical limit by probing
the curvature of the entropy surface. There still exists open
problems.
One of them is the scaling of a first order transition: in
Fig.~\ref{fig:SEx} in a canonical sense the phase transition occurs
from $X_A$ to $X_D$ (if $X$ is the energy $E$ then $X_D-X_A$ is the
transition latent heat), but for if one uses the ME definition the
transition occurs only from $X_B$ to $X_C$. At the thermodynamical
limit, if it exists, such discrepancy should disappear. In another
context (a model of a first order liquid gas transition of
finite--size Sodium clusters) we could show that
$\frac{X_C-X_B}{X_D-X_A}\to1$ when the system size goes to infinity.

The ME definition of phase transitions offers a richer view of
physical systems and phase transitions. Again on Fig.~\ref{fig:SEx},
CE is not sensitive to everything that could happen under the Maxwell
line: there is always one transition. On the contrary, in ME there
could be many phase transitions between $X_A$ and $X_D$. For example
if there is a small positive curvature bump between $X_B$ and $X_C$
there would be two transitions in the microcanonical sense but still
one in the canonical sense.



\acknowledgments We are gratefull to V. Laliena, D. Valls--Gabaud and
P.--H. Chavanis for useful comments and criticisms. We also thank E.
Votyakov for discussions and technical help.

\appendix

\section{}
\label{sec:appendix}

Let us compute $\langle\vp_k\rangle_{\vq_k}$ the average momentum of particle
$k$ at fixed position (for simplicity we will set $k=1$). The
$\alpha$--component of $\langle\vp_1\rangle_{\vq_1}$ is
\begin{eqnarray}
  \nonumber
    \langle p_1^\alpha\rangle_{\vq_1}&=&\frac{\int \left(\prod_id\vp_i \prod_{i=2}^Nd\vq_i\right) p_1^\alpha
    \delta\left(E-{\cal H}\right)
    \delta^2\left(\sum_i\vp_i\right)
    \delta\left(\sum_i\vq_i\times\vp_i-L\right)
    \delta^2\left(\sum_i\vq_i\right)}{\int \left(\prod_id\vp_i \prod_{i=2}^Nd\vq_i\right)
    \delta\left(E-{\cal H}\right)
    \delta^2\left(\sum_i\vp_i\right)
    \delta\left(\sum_i\vq_i\times\vp_i-L\right)
    \delta^2\left(\sum_i\vq_i\right)} \\
  \label{eq:32}
  & = & \frac{\int \left(\prod_{i=2}^Nd\vq_i\right) {\cal P}_1^\alpha
    \delta^2\left(\sum_i\vq_i\right)}{\int \left(\prod_{i=2}^Nd\vq_i\right)
    W({\bf r})
    \delta^2\left(\sum_i\vq_i\right)},
\end{eqnarray}
where ${\cal P}_1^\alpha=\int \left(\prod_id\vp_i\right) p_1^\alpha \delta\left(E-{\cal
    H}\right) \delta^2\left(\sum_i\vp_i\right)
\delta\left(\sum_i\vq_i\times\vp_i-L\right)$, and $W({\bf r})$ is the
microcanonical weight at fixed spatial configuration ${\bf r}$, its
value is~$W(E,L,{\bf r})={\cal C}\frac{1}{\sqrt{I}}E_r^{N-5/2}$
(see~(\ref{eq:5})). The outline of the derivation of ${\cal P}_1^\alpha$ is
the same as in~\cite{laliena99:_effec} for $W$, and we get after some
algebra
\begin{equation}
  \label{eq:33}
  {\cal P}_1^\alpha={\cal C}Lm_1I^{-3/2}\sum_{\delta=1}^2r_1^\delta\epsilon_{\delta\alpha} \:
  E_r^{N-5/2},
\end{equation}
where ${\bf \epsilon}$ is the antisymmetric tensor of rank 2. Using
(\ref{eq:33}) in (\ref{eq:32}) we get finally
\begin{eqnarray}
  \nonumber
  \langle p_1^\alpha\rangle_{\vq_1}&=&\frac{\int \left(\prod_{i=2}^Nd\vq_i\right) 
    {\cal C}Lm_1I^{-3/2}\sum_{\delta=1}^2r_1^\delta\epsilon_{\delta\alpha} \:
    E_r^{N-5/2}
    \delta^2\left(\sum_i\vq_i\right)}{\int \left(\prod_{i=2}^Nd\vq_i\right)
    I^{-1/2}\sum_{\delta=1}^2r_1^\delta\epsilon_{\delta\alpha} \:
    E_r^{N-5/2}    \delta^2\left(\sum_i\vq_i\right)} \\
  \label{eq:34}
  & = & Lm_1\langle I^{-1}\rangle_{{\bf r}_k}\sum_\delta r_1^\delta\epsilon_{\delta\alpha}.
\end{eqnarray}
Finally
\begin{equation}
  \label{eq:35}
  \langle\vp_1\rangle_{\bmr_1}=Lm_1\langle I^{-1}\rangle_{{\bf r}_k}\sum_{\delta,\alpha} r_1^\delta\epsilon_{\delta\alpha}{\bf e_\alpha},
\end{equation}
where ${\bf e_\alpha}$ is the $\alpha$--component unit vector.

$\langle\vp_k^2\rangle_{\vq_k}$ can be derived in a similar way, and we get
\begin{eqnarray}
  \label{eq:36}
  \nonumber
  \langle \vp_k^2\rangle_{\vq_k} &=& \frac{2m_k}{(N-5/2)\langle Er^{-1}\rangle_{\vq_k}}
  \lb 1-\frac{m_k}{M} \rb\\
    & &- \frac{m_kI_k}{I(N-5/2)\langle Er^{-1}\rangle_{\vq_k}} +I_kL^2m_k\langle I^{-2}\rangle_{\vq_k}.
\end{eqnarray}


\begin{thebibliography}{10}

\bibitem{antonov62}
V. Antonov, Vestn, Leningr. Gros. Univ. {\bf 7},  135  (1962).

\bibitem{padmanabhan90:_statis}
T. Padmanabhan, Physics Report  {\bf 188},  285  (1990).

\bibitem{1967MNRAS.136..101L}
D. {Lynden--Bell}, Mon. Not. R. Astron. Soc.  {\bf 136},  101+  (1967).

\bibitem{1968MNRAS.138..495L}
D. {Lynden--Bell} and R. {Wood}, Mon. Not. R. Astron. Soc.  {\bf 138},  495+  (1968).

\bibitem{hertel71:_solub_system_negat_specif_heat}
P. Hertel and W. Thirring, Annals of Physics {\bf 63},  520  (1971).

\bibitem{thirring70:_system_negat_specif_heat}
W. Thirring, Z. Phys.  {\bf 235},  339  (1970).

\bibitem{saslaw85:_gravit}
W.~C. Saslaw, {\em Gravitational physics of stellar and galactic systems}
  (Cambridge University Press, Cambridge, 1985).

\bibitem{1999MNRAS.309..481L}
G.~B. {Lima Neto}, D. {Gerbal}, and I. {M{\'a}rquez}, Mon. Not. R. Astron. Soc.  {\bf 309},  481
   (1999).

\bibitem{lima00}
A.~R. Lima, P.~M.~C. de~Oliveira, and T.~J.~P. Penna, J. Stat. Phys. {\bf 99},
  691  (2000).

\bibitem{binney87:_galac}
J. Binney and S. Tremaine, {\em Galactic dynamics} (Princeton series in
  astrophysics, Princeton, NJ, 1987).

\bibitem{lagoute96:_rotat_1}
C. Lagoute and P.-Y. Longaretti, Astron. Astrophys. {\bf 308},  441  (1996).

\bibitem{lagoute96:_rotat_2}
C. Lagoute and P.-Y. Longaretti, Astron. Astrophys. {\bf 308},  453  (1996).

\bibitem{horwitz77:_steep}
G. Horwitz and J. Katz, Astron. J. {\bf 211},  226  (1977).

\bibitem{klinko00:_mean_field_theor_spher_gravit_system}
P.~J. Klinko and B.~N. Miller, Mean Field Theory of Spherical Gravitating
  Systems, e--print arXiv.org/cond-mat/0007065, 2000.

\bibitem{lynden-bell00:_rotat_statis}
D. Lynden-Bell, Rotation, Statistical Dynamics and Kinematics of Globular
  Clusters, e--print arXiv.org/astro-ph/0007116, 2000.

\bibitem{combes98:_fract_struc_driven_self_gravit}
F. Combes,  in {\em Chaotic Dynamics of Gravitational Systems}, Les Arcs 2000,
  edited by C. Mechanics (1998).

\bibitem{vega:_fract_struc_scalin_laws_univer}
H.~J. de~Vega, N. S{\'a}nchez, and F. Combes, Fractal Structures and Scaling
  Laws in the Universe: Statistical Mechanics of the Self--Gravitating Gas,
  Invited paper to the special issue of the `Journal of Chaos, Solitons and
  Fractals': `Superstrings, M, F, S...theory', M. S El Naschie and C. Castro,
  Editors.

\bibitem{bate98:_collap_molec_cloud_core_stell_densit}
M.~R. Bate, Astron. J. Lett. {\bf 518},  L95  (1998).

\bibitem{klein98:_gravit_collap_fragm_molec_cloud}
R.~I. Klein, R.~T. Fisher, C.~F. McKee, and J.~K. Truelove,  in {\em
  Proceedings of Numerical Astrophysics 1998} (1998).

\bibitem{klessen97:_fragm_molec_cloud}
R.~S. Klessen and A. Burkert,  in {\em Riken Symposium on Supercomputing}
  (1997).

\bibitem{1996MNRAS.280.1190B0}
A. {Burkert} and P. {Bodenheimer}, Mon. Not. R. Astron. Soc.  {\bf 280},  1190  (1996).

\bibitem{whitworth96:_star}
A.~P. Whitworth, N.~F. A.~S.~Bhattal, and S.~J. Watkins, Mon. Not. R. Astron. Soc.  {\bf 283},
  1061  (1996).

\bibitem{white96:_violen_relax_hierar_clust}
S.~D.~M. White,  in {\em Gravitational Dynamics}, edited by E.~T. R.~Terlevich,
  O.~Lahav (Cambridge Univ. Press, Cambridge, 1996).

\bibitem{1990ApJ...363..197C}
D.~M. {Christodoulou} and J.~E. {Tohline}, Astron. J. {\bf 363},  197  (1990).

\bibitem{laliena99:_effec}
V. Laliena, Phys. Rev. E {\bf 59},  4786  (1999).

\bibitem{yawn97:_ergod}
K.~R. Yawn and B.~N. Miller, Phys. Rev. E {\bf 56},  2429  (1997).

\bibitem{reidl93:_gravit}
C.~J. {Reidl, Jr.} and M.~B. N., Phys. Rev. E {\bf 48},  4250  (1993).

\bibitem{sommer-larsen97:_struc_isoth_self_gas_spher_soften_gravit}
J. Sommer-Larsen, H. Vedel, and U. Hellsten, Astron. J. {\bf 500},  610  (1997).

\bibitem{follana00:_therm_self_gravit_system_soften_poten}
E. Follana and V. Laliena, Phys. Rev. E {\bf 61},  6270  (2000).

\bibitem{chavanis98:_system}
P.-H. Chavanis, Phys. Rev. E {\bf 58},  R1199  (1998).

\bibitem{lee52:_statis_theor_II}
T.~D. Lee and C.~N. Yang, Phys. Rev. {\bf 87},  410  (1952).

\bibitem{gross00:_phase_trans_small}
D.~H.~E. Gross and E. Votyakov, Eur. Phys. J. B {\bf 15},  115  (2000).

\bibitem{gross97:_microc}
D.~H.~E. Gross, Physics Report {\bf 279},  119  (1997).

\bibitem{plummer11}
H.~C. Plummer, Mon. Not. R. Astron. Soc.  {\bf 71},  460  (1911).

\bibitem{yepes97:_cosmol_numer} G. Yepes, in {\em From quantum
    fluctuations to cosmological structures}, Vol.~126 of {\em ASP
    Conferences}, edited by D. Valls-Gabaud, M. Hendry, P.  Molaro,
  and K. Chamcham (Astronomical Society of the Pacific, Provo, Utah
  USA, 1997), p.\ 279.

\bibitem{calvo98:_monte_carlo}
F. Calvo and P. Labastie, Eur. Phys. J. D {\bf 3},  229  (1998).

\bibitem{b.89:_physiq_statis}
B. Diu, C. Guthmann, D. Lederer, and B. Roulet, {\em Physique Statistique}
  (Hermann, 293 rue Lecourbe, 75015 Paris, France, 1989).

\bibitem{kunz94:_multip}
R.~E. Kunz and R. Berry, Phys. Rev. E {\bf 49},  1895  (1994).

\bibitem{gross95:_statit}
D.~H.~E. Gross and P.~A. Hervieux, Z. Phys. D  {\bf 33},  295  (1995).

\bibitem{posch90:_dynam}
H.~A. Posch, H. Narnhofer, and W. Thirring, Phys. Rev. A {\bf 42},  1880  (1990).

\bibitem{gross97:_fragm_i}
D.~H.~E. Gross, M.~E. Madjet, and O. Schapiro, Z.Phys.D {\bf 39},  75  (1997).

\bibitem{landau94:_physiq}
L. Landau and E. Lifchitz,  in {\em Physique statistique} (Mir--Ellipses,
  Moskow, 1994), Chap.~II.26.

\bibitem{torcini99:_equil_n}
A. Torcini and M. Antoni, Phys. Rev. E {\bf 59},  2746  (1999).

\bibitem{lee93:_new_monte_carlo}
J. Lee, Phys. Rev. Lett. {\bf 71},  211  (1993).

\bibitem{berg93:_simul}
B.~A. Berg and U. Hansmann, Phys. Rev. B {\bf 47},  497  (1993).

\bibitem{ferrenberg89:_optim_monte_carlo}
A.~M. Ferrenberg and R.~H. Swendsen, Phys. Rev. Lett. {\bf 63},  1195  (1989).

\bibitem{smith96:_free_energ_monte_carlo}
G.~R. Smith, Ph.D. thesis, University of Edinburgh, 1996.

\bibitem{fliegans01:_phase}
O. Fliegans, Ph.D. thesis, Freie Universit{{\"a}}t Berlin, 2001.

\bibitem{convex}
A set ${\cal C}$ of ${\Bbb R}^{n}$ is said to be {\em convex\/} if for any
  couple of points $M$ and $N$ ($M\neq N$) of ${\cal C}$, the segment $[M,N]$
  is in ${\cal C}$.

\bibitem{vega00}
H.~J. de~Vega and N. S{\'a}nchez, Physics Letter B {\bf 490},  180  (2000).

\bibitem{sota00:_origin_fract_distr_self_gravit}
Y. Sota, O. Iguchi, M. Morikawa, T. Tatekawa, and K. ichi Maeda, The Origin of
  Fractal Distribution in Self--Gravitating Virialized System and
  Self--Organized Criticality, e--print arXiv.org/astro-ph/0009412, 2000.

\bibitem{semelin99:_physiq}
B. Semelin, Ph.D. thesis, Universit{\'e} Paris VI, 1999.

\bibitem{ehrenfest12a}
P. Ehrenfest and T. Ehrenfest, {\em The Conceptual Foundation of the
  Statistical Approach in Mechanics} (Cornell University Press, Ithaca NY,
  1959), pp.\ 20--22.

\bibitem{sato00:_irrev}
K. Sato, K. Sekimoto, T. Hondou, and F. Takagi, Irreversibility resulting from
  contact with a heat bath caused by the finiteness of the system, e--print
  arXiv.org/cond-mat/0008393, 2000.
  
\bibitem{sekimoto00:_carnot_cycle_small_system} K. Sekimoto, F.
  Takagi, and T. Hondou, The Carnot Cycle for Small Systems:
  Irreversibility and the Cost of Operations, Phys. Rev. E {\bf 62},
  7759 (2000), e--print arXiv.org/astro-ph/0003064.

\bibitem{2000NewA....4..601H}
A. {Hujeirat}, P. {Myers}, M. {Camenzind}, and A. {Burkert}, New Astronomy {\bf
  4},  601  (2000).

\bibitem{galli00:_singul_isoth_disks_format_multip_stars} D. Galli,
  F.~H. Shu, G. Laughlin, and S. Lizano, in {\em Stellar Clusters and
    Associations}, Vol.~198 of {\em ASP Conference Series}, edited by
  T.  Montmerle and P. Andr{\'e} (ASP, Provo, Utah USA, 2000).

\bibitem{2000A&A...356.1089C}
P.-H. {Chavanis}, Astron. Astrophys. {\bf 356},  1089  (2000).

\end{thebibliography}
\end{document}